\documentclass[aps, showpacs, showkeys,nofootinbib,floatfix]{revtex4}

\usepackage{amssymb}
\usepackage{amsmath}
\usepackage{graphicx}

\begin{document}

\title{Testing cosmic opacity from SNe Ia and Hubble parameter through three cosmological-model-independent methods}

\author{Kai Liao\footnote{liaokai@mail.bnu.edu.cn}, Zhengxiang Li, Jing Ming, Zong-Hong Zhu}

\affiliation{Department of Astronomy, Beijing Normal University,
Beijing 100875, China}

\begin{abstract}
We use the newly published 28 observational Hubble parameter data ($H(z)$) and current largest SNe Ia samples (Union2.1)
to test whether the universe is transparent. Three cosmological-model-independent methods (nearby SNe Ia method,
interpolation method and smoothing method) are proposed through comparing opacity-free distance modulus from Hubble parameter data
and opacity-dependent distance modulus from SNe Ia . Two parameterizations, $\tau(z)=2\epsilon z$ and $\tau(z)=(1+z)^{2\epsilon}-1$
are adopted for the optical depth associated to the cosmic absorption. We find that the results are not sensitive to
the methods and parameterizations. Our results support a transparent universe.

\end{abstract}


\maketitle

\section{Introduction}
The unexpected dimming of Type Ia supernova (SNe Ia) is thought to be the evidence of acceleration of the universe \cite{acceleration}.
In the frame of General Relativity (GR), the most famous explanation is the existence of dark energy with a negative pressure \cite{dark energy}.
However, there are some issues on this plausible mechanism for observed SNe Ia dimming. The photon number
conservation may be deviated. For example, it is due to the dust in our galaxy and oscillation of photons propagating
in extragalactic magnetic fields into very light axions. These absorption, scattering or axion-photon mixing may
lead to dimming \cite{issue}. Other mechanisms are widely proposed including modified gravity \cite{modified}, dissipative processes \cite{dissipative},
evolutionary effects in SNe Ia events \cite{evolution}, violation of cosmological principle \cite{LTB} and so on.
On the other hand, the deviation of photon number conservation is related to the correction of Tolman test \cite{Tolman} which is equivalent to
measurements of the well-known distance-duality relation (DDR) \cite{DDR}
\begin{equation}
 \frac{D_{\scriptstyle L}}{D_{\scriptstyle A}}{(1+z)}^{-2}=1\;,
 \label{rec}
\end{equation}
where $D_{\scriptstyle L}$ is luminosity distance, $D_{\scriptstyle A}$ is angular diameter distance and $z$ is redshift.
The DDR is in fact a particular form of presenting the general theorem proved by Etherington known as the "reciprocity law" or "Etherington reciprocity theorem".
DDR holds for general metric theories of gravity in any cosmic background and it is valid for any cosmological
models based on the Riemannian geometry. It is independent of gravity equation and the universe components.
However, DDR may be not valid in the case that photons do not travel along null geodesics or the cosmic opacity exists. Many efforts have been done to test DDR though astronomical observations \cite{DDRobs}. Usually, they assume the form $\frac{D_{\scriptstyle L}}{D_{\scriptstyle A}}{(1+z)}^{-2}=\eta(z)\;$, where $\eta(z)=1+\eta_0z$ or $\eta(z)=1+\eta_0\frac{z}{1+z}$.
Compared to conservation of photon number, the assumptions that the mathematical tool used to describe the space-time of universe is Riemannian geometry and photon travels along null geodesic are more fundamental and unassailable, thus the deviation of DDR most possibly indicates cosmic absorption.
In this case, the flux received by the observer will be reduced by a factor $e^{-\tau(z)}$, and observed luminosity distance can be obtained by \cite{depth}
\begin{equation}
 D_{L,obs}=D_{L.true}e^{\tau(z)/2},
 \label{rec}
\end{equation}
where $\tau(z)$ is the optical depth related to the cosmic absorption. The relation between $\tau(z)$ and $\eta(z)$ is
$e^{\tau(z)/2}=\eta(z)$ \cite{depth2}.
Following this assumption, Avgoustidis et al. \cite{Avgoustidis} studied the difference between SNe Ia observations and Hubble
parameter data. $H(z)$ data are mainly obtained through the measurements of differential ages of
red-envelope galaxies known as "differential age method". The aging of stars can be regarded as an indicator
of the aging of the universe. The spectra of stars can be converted to the information of their ages, as the
evolutions of stars are well known. Since the stars cannot be observed one by one at cosmological scales, people
usually take the spectra of galaxies which contain relatively uniform star population.
Moreover, $H(z)$ data can be obtained from the BAO scale as a standard ruler
in the radial direction known as "Peak Method". These methods are apparently independent of galaxy luminosity so
that it will not be affected by cosmic opacity. However, SNe Ia observations are affected by many sources of opacity,
such as the hosting galaxy, intervening galaxies, intergalactic medium, the Milky Way and exotic physics which affect photon conservation. Under the assumption
$D_L=D_A(1+z)^{2+\epsilon}$, they investigated the cosmic opacity by confronting the standard luminosity distance in spatially flat $\Lambda$CDM model with the observed one from SNe Ia observations. Combining with the $H(z)$ data, which is not affected by transparency but yields constraints on $\Omega_m$, and marginalizing over all other parameters except $\epsilon$, they got $\epsilon=-0.04_{-0.07}^{+0.08}$ (2$\sigma$). Noticing that their method depends on cosmological models, Holanda et al. \cite{Holanda} further proposed a model-independent
estimate of $D_L$ which are obtained from a numerical integration of $H(z)$ data. They also explored the influence of
different SNe Ia light-curve fitters (SALT2 and MLCS2K2) and found a significant conflict.
Based on Holanda et al., we proposed three model-independent methods that are different from theirs to explore the cosmic opacity. They firstly got the
luminosity distances from $H(z)$ data (12 data) at corresponding redshifts (at $H(z)$ data) and gave a polynomial fit based on these 12 luminosity distances data with
their errors, then calculated the values at the redshifts corresponding to SNe Ia through this polynomial fit curve. On the contrary, we use SNe Ia data
to get the luminosity distances at the redshifts corresponding to $H(z)$ data through interpolation method,
smoothing method and nearby SNe Ia method. Our data sets contains 28 available $H(z)$ data and the largest SNe Ia samples Union2.1 \cite{Union2.1}.

The Letter is organized as follows: In Section 2, we introduce the method of obtaining luminosity distance from $H(z)$ data.
In Section 3, we give the three methods that can convert SNe Ia luminosity distances to the luminosity distances at the
redshifts corresponding to $H(z)$ data. In Section 4, the results are performed. Finally, we make a conclusion in
Section 5.

\section{Luminosity distance from observational Hubble parameter data}
In this section, we introduce the method proposed by Holanda et al. \cite{Holanda}. The expression of the Hubble parameter can be
written in this form
\begin{equation}
H(z)=-\frac{1}{1+z}\frac{dz}{dt},
\end{equation}
which depends on the differential age as a function of redshift. Based on Jimenez et al. \cite{Jimenez}, Simon et al. \cite{Simon}
used the age of evolving galaxies and got nine $H(z)$ data. Stern et al. \cite{Stern} revised these data at 11 redshifts
from the differential ages of red-envelope galaxies. Gazta\~{n}aga et al. \cite{Gaztanaga} took the BAO scale as a standard
ruler in the radial direction, obtained two data. Recently, Moresco et al. \cite{Moresco}
obtained 8 data from the differential spectroscopic evolution of early-type galaxies as
a function of redshift. Blake et al. \cite{Blake} obtained 3 data through combining measurements
of the baryon acoustic peak and Alcock-Paczynski distortion from galaxy clustering in the WiggleZ Dark Energy Survey. Zhang et al. \cite{zhang} obtained another 4 data. Totally, we have 28 available data summarized in Table 1.
Following Holanda et al. \cite{Holanda}, we transform these 28 $H(z)$ data into luminosity distance. Using a usual
simple trapezoidal rule, the comoving distance can be calculated by
\begin{equation}
r=c \int_0^z{dz^\prime \over H(z^\prime)}\approx {c\over 2}\sum_{i=1}^{n} (z_{i+1}-z_i)\left[ {1\over H(z_{i+1})}+{1\over H(z_i)} \right].
\end{equation}
Since we use much more data than Holanda et al. (they used only 12 data), this trapezoidal rule will work much better. In Fig. 1, we show the relative error with respect to the data number at a characteristic redshift $z=1$. We assume a standard $\Lambda$CDM model with $\Omega_m=0.3,\Omega_\Lambda=0.7$, and divide $z=1$ into
different numbers of intervals averagely, then calculate the relative errors according to
Eq. (4). We find that the relative errors decrease remarkably when the numbers of intervals increase from 12 to 28.
With the standard error propagation formula, the error associated to the $i^{th}$ bin is given by
\begin{equation}
s_i={c\over 2}(z_{i+1}-z_i)\left({\sigma_{H_{i+1}}^2\over H_{i+1}^4} + {\sigma_{H_{{i}}}^2\over H_{i}^4}\right)^{1/2}\;,
\end{equation}
where $\sigma_{H_{i}}$ is the error of $H(z)$ data. The error corresponding to certain redshift is the sum of $s_i$. The Hubble constant $H_0=73.8 \pm 2.4$ km/s/Mpc \cite{h0} is used in our study. The 28 luminosity distance data from $H(z)$ data are shown in Fig. 2, as well
as the $D_L$ from Union2.1 SNe Ia samples.

\begin{figure}[!htbp]
\includegraphics[width=12cm]{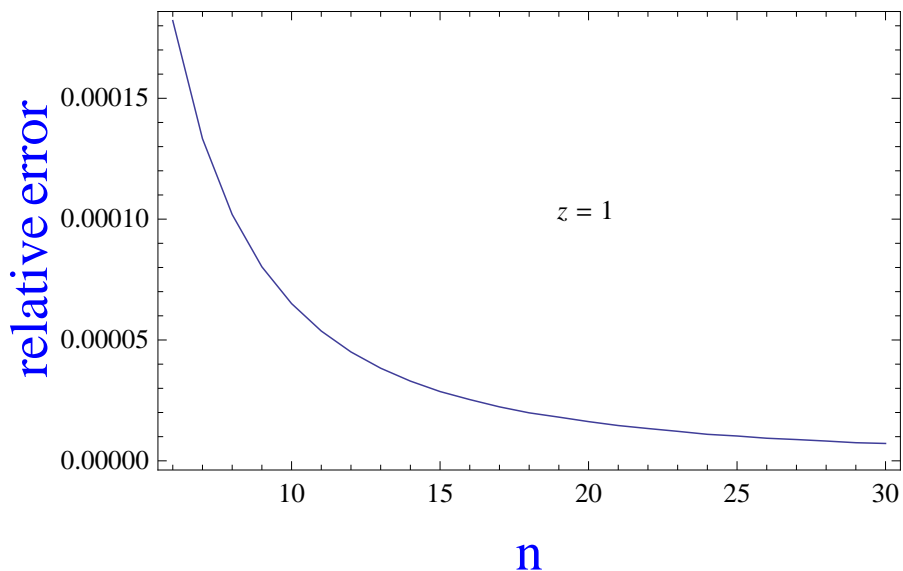}
\caption{Relative error with respect to the data number at $z=1$. $\Omega_m=0.3$, $\Omega_\Lambda=0.7$.
 }
\end{figure}

\begin{figure}[!htbp]
\includegraphics[width=12cm]{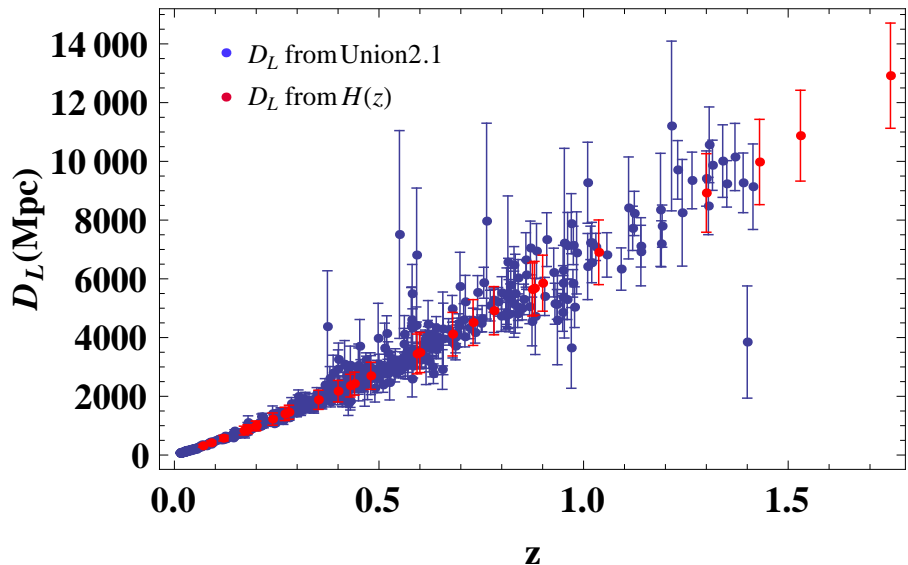}
\caption{$D_L(z)$ obtained from measurements of the Hubble parameter data
 and the Union2.1 samples, respectively.
 }\label{DL}
\end{figure}

\section{Dealing with SNe Ia samples}
In this section, we introduce three methods through which we can obtain the luminosity
distance of one certain SNe Ia point at the same redshift of the corresponding $H(z)$ data.
\subsection{nearby SNe Ia method}
Since the SNe Ia Samples is much larger than $H(z)$ data, the nearby SNe Ia can be substituted
for the one at the redshift corresponding to $H(z)$ data. Points $z_{SNe~Ia}-z_{H}$ are centered
around the line $\Delta z=0$, as shown in Fig. 3 which plots the subtractions of redshifts between
$H(z)$ data and the associated SNe Ia. Similar with the DDR test \cite{DDRobs}, we have to choose a
criterion based on the data. Our selection criterion is
$\Delta z=\left|z_{{H}}-z_{{SNe~Ia}}\right|<0.003$. This selection criteria can be
satisfied for most of the $H(z)$ data except for the points at $z=0.9$ and $z=1.037$ ($z=1.43, 1.53, 1.75$ are obviously ruled out) and small enough to reduce the systematic errors and guarantee the accuracy.

\begin{figure}[!htbp]
\includegraphics[width=12cm]{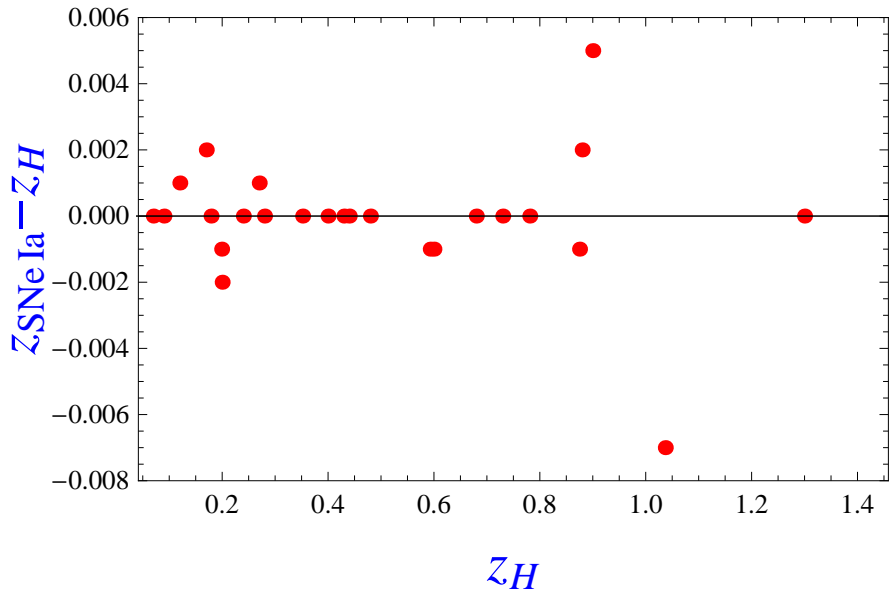}
\caption{Subtractions of redshifts between $H(z)$ data and the associated SNe Ia.
 }
\end{figure}

\subsection{interpolation method}
In order to avoid any bias brought by redshift incoincidence between $H(z)$ data and SNe Ia, as well as to ensure the integrity of the $H(z)$ data, we can use the nearby SNe Ia points to obtain
the luminosity distance of SNe Ia point at the same redshift of the corresponding $H(z)$ data. This situation is similar with the cosmology-independent calibration of GRB relations directly
from SNe Ia \cite{Interpolation}. When the linear interpolation is used, the distance modulus
and the error can be calculated by
\begin{equation}
\mu = \mu_i+(\mu_{i+1}-\mu_i)[(z-z_{i})/(z_{i+1}-z_i)],
\end{equation}
\begin{equation}
\sigma_{\mu}=([(z_{i+1}-z)/(z_{i+1}-z_i)]^2\sigma_{\mu,i}^2+[(z-z_{i})/(z_{i+1}-z_i)]^2\sigma_{\mu,i+1}^2)^{1/2},
\end{equation}
where subscripts $i$ and $i+1$ stand for the nearby data.
We use the same method as used in \cite{Schaefer} to obtain the best estimate for each SNe Ia which
is weighted average of all available distance moduli at the same redshift. The
derived distance modulus for each SNe Ia is

\begin{equation}
\mu = (\sum_i \mu_{\rm i} / \sigma_{\mu_{\rm i}}^2)/(\sum_i
\sigma_{\mu_{\rm i}}^{-2}),
\end{equation}
with its uncertainty $ \sigma_{\mu} = (\sum_i \sigma_{\mu_{\rm
i}}^{-2})^{-1/2}$.

\subsection{smoothing method}
We introduce a non-parametric method of smoothing supernova data over redshift using a Gaussian kernel in order
to reconstruct luminosity distance \cite{smooth}. This procedure was initially used in the analysis of the cosmic large
scale structure \cite{structure}. Through this model-independent method, we can extract information of various cosmological parameters, such
as Hubble parameter, the dark energy equation of state, the matter density. Wu and Yu \cite{generalized} generalized this method to
eliminate the impact of $H_0$ and obtained the evolutionary curve of luminosity distance using SNe Ia
Constitution set and Union2 set. In this Letter, we follow this generalized method to get the luminosity distance curve
using Union2.1 set. Firstly, we obtain the variable $\ln f(z)=\ln D_L(z)-\ln h$ through iterative method

\begin{equation}
\ln f(z)_n^s=\ln\ f(z)_{n-1}^s+
N(z) \sum_i \left [ \ln f^{obs}(z_i)-\ln\ f(z)_{n-1}^s \right]
{\large \times} \ {\rm exp} \left [- \frac{\ln^2 \left
( \frac{1+z}{1+z_i} \right ) }{2 \Delta^2} \right ],
\end{equation}
where the reduced Hubble constant $h=H_0/100$ and the normalization factor
\begin{equation}
N(z)^{-1}=\sum_i {\rm exp} \left
[- \frac{\ln^2 \left ( \frac{1+z}{1+z_i} \right ) }{2 \Delta^2}
\right ].
\end{equation}
The value of parameter $\Delta$ was discussed by Shafieloo et al. \cite{smooth}. The results are not sensitive
to $\Delta$. Following Shafieloo et al. \cite{smooth} and Wu and Yu \cite{generalized}, $\Delta=0.6$ is used in this Letter.
Eq. (9) can give the smoothed luminosity distance at any redshift z after n iterations.
When $n=1$

\begin{equation}
\begin{split}
\ln f(z)_1^s=\ln\ f(z)_0^s+
N(z) \sum_i \left [ \ln f^{obs}(z_i)-\ln\ f(z)_0^s \right]
{\large \times} \ {\rm exp} \left [- \frac{\ln^2 \left
( \frac{1+z}{1+z_i} \right ) }{2 \Delta^2} \right ]\\
=\ln\ D_L(z)_0^s+
N(z) \sum_i \left [ \ln f^{obs}(z_i)-\ln\ D_L(z)_0^s \right]
{\large \times} \ {\rm exp} \left [- \frac{\ln^2 \left
( \frac{1+z}{1+z_i} \right ) }{2 \Delta^2} \right ],
\end{split}
\end{equation}
where$D_L(z)_0^s$ depends on the suggested background cosmological model. Following Wu an Yu \cite{generalized},
we adopt $\omega$CDM model with $\omega=-0.9$ and $\Omega_{m0}=0.28$ as the background model.
The relation between $\ln f^{obs}(z_i)$ and $\ln D^{obs}_L(z_i)$ which is obtained from observed
SNe Ia is

\begin{equation}
\ln f^{obs}(z_i)={\ln 10\over 5} [\mu^{obs}(z_i)-42.38]=\ln D^{obs}_L(z_i)-\ln h,
\end{equation}
$\mu^{obs}(z_i)$ is the observed distance modulus from SNe Ia. In order to know whether we
get a best-fit value after some iterations, we calculate, after each iteration, $\chi^2_s$

\begin{equation}
\chi^2_{s,n}=\sum_i {(\mu(z_i)_n-\mu^{obs}(z_i))^2 \over
\sigma^2_{\mu_{obs,i}}}\;.
\end{equation}
The best-fit result is corresponding to the minimum value of $\chi^2_{s,n}$. The 1 $\sigma$ corresponds to $\Delta \chi^2_s=1$. The results For Union2.1 samples are shown in Fig. 4 and Fig. 5, the minimum value $n=32$.

\begin{figure}[!htbp]
\includegraphics[width=12cm]{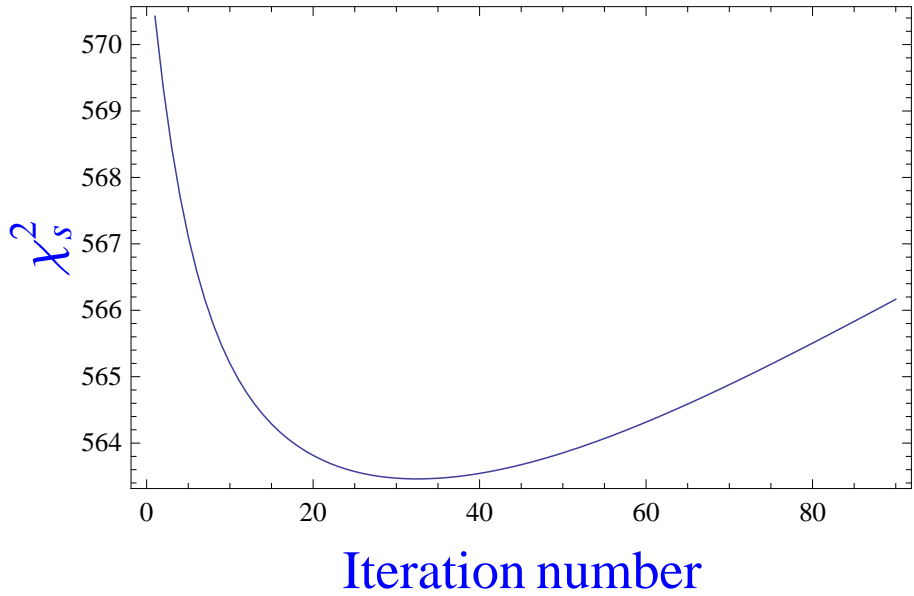}
\caption{Computed $\chi_s^2$ for the reconstructed results at each iteration.
 }
\end{figure}

\begin{figure}[!htbp]
\includegraphics[width=12cm]{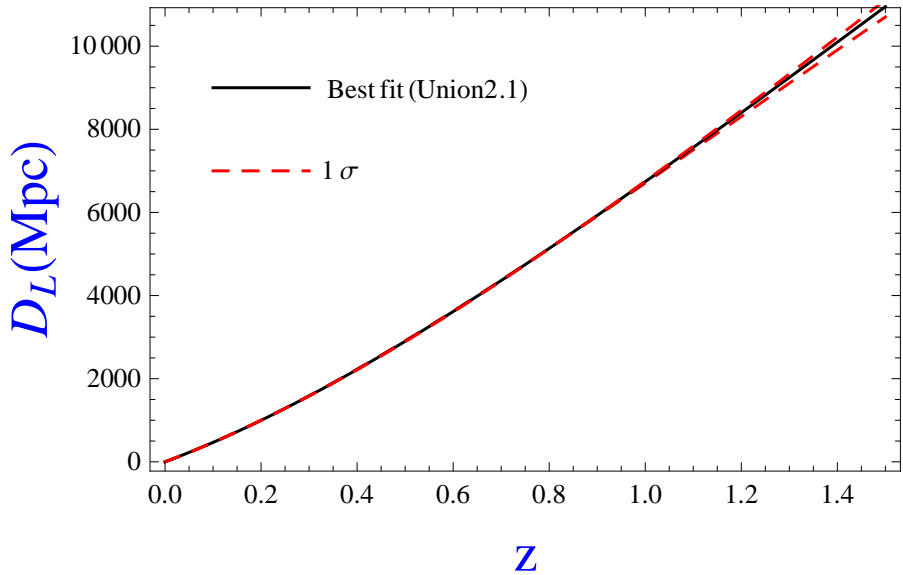}
\caption{Smoothed luminosity distance for Union2.1 samples.
 }
\end{figure}

\section{Constraints on cosmic opacity}
The observed distance modulus can be expressed as \cite{depth}
\begin{equation}
\mu_{obs}(z)=\mu_{true}(z)+2.5[\log_{10} e] \tau(z) \, .
\end{equation}
To examine the sensitivity of test results on the parametric form, we adopt two parameterizations, $\tau(z)=2\epsilon z$ and $\tau(z)=(1+z)^{2\epsilon}-1$ which are not strongly wavelength dependent on the optical band \cite{form}. $\epsilon$ here describes the cosmic opacity. The former one is linear and it can be derived from the
DDR parameterization $D_L=D_A(1+z)^{2+\epsilon}$ for small $\epsilon$ and redshift. To constrain the value of $\epsilon$,
we use the usual maximum likelihood method of $\chi^2$ fitting

\begin{equation}
\chi^{2} =  \sum_{i}\frac{(\mu_{obs}(i) - \mu_{true}(i)-2.5[\log_{10} e] \tau(i))^2}
{\sigma^2_{\mu(obs)}+ \sigma^2_{\mu(true)}},
\end{equation}
the subscript i stands for the data at the redshifts corresponding to $H(z)$ data. Our results are shown in Fig. 6, Fig. 7 and Fig. 8, respectively. For smoothing method, we consider two cases: $i_{max}=25$ and $i_{max}=28$ (containing the data at $z=1.43, 1.53, 1.75$). From the likelihood of $\epsilon$ using different methods, we can see current SNe Ia samples and $H(z)$ data support a transparent universe. These results are slightly different from Holanda et al. \cite{Holanda} while
their results seem a little prone to a non-transparent universe especially with MLCS2K2 compilation.

\begin{figure}[!htbp]
\includegraphics[width=12cm]{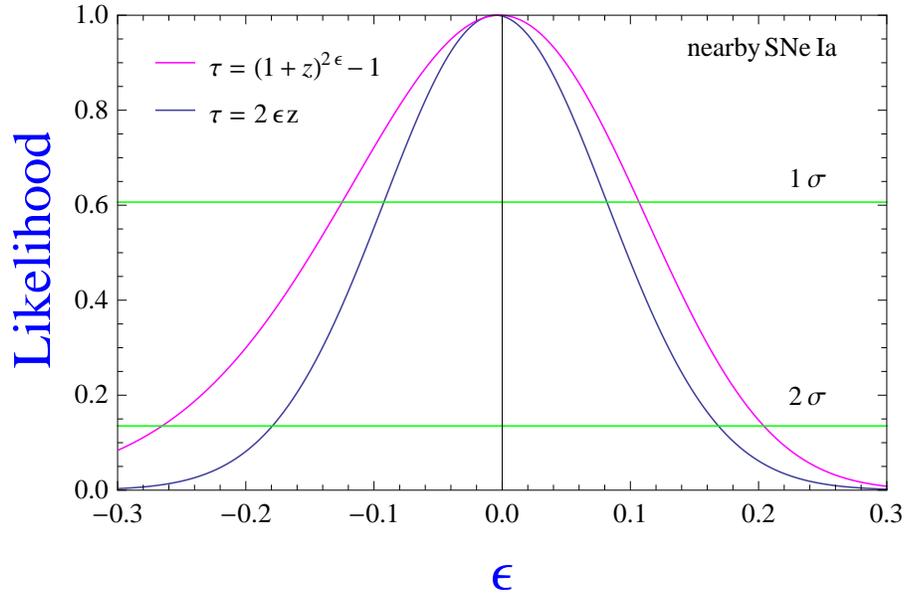}
\caption{Likelihood function for $\epsilon$ using nearby SNe Ia method.
 }
\end{figure}

\begin{figure}[!htbp]
\includegraphics[width=12cm]{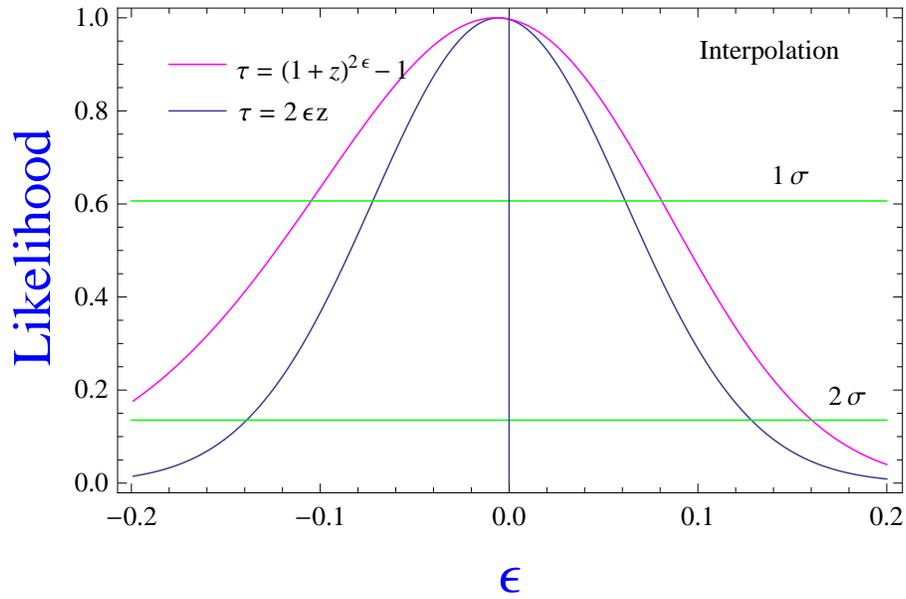}
\caption{Likelihood function for $\epsilon$ using interpolation method.
 }
\end{figure}

\begin{figure}[!htbp]
\includegraphics[width=12cm]{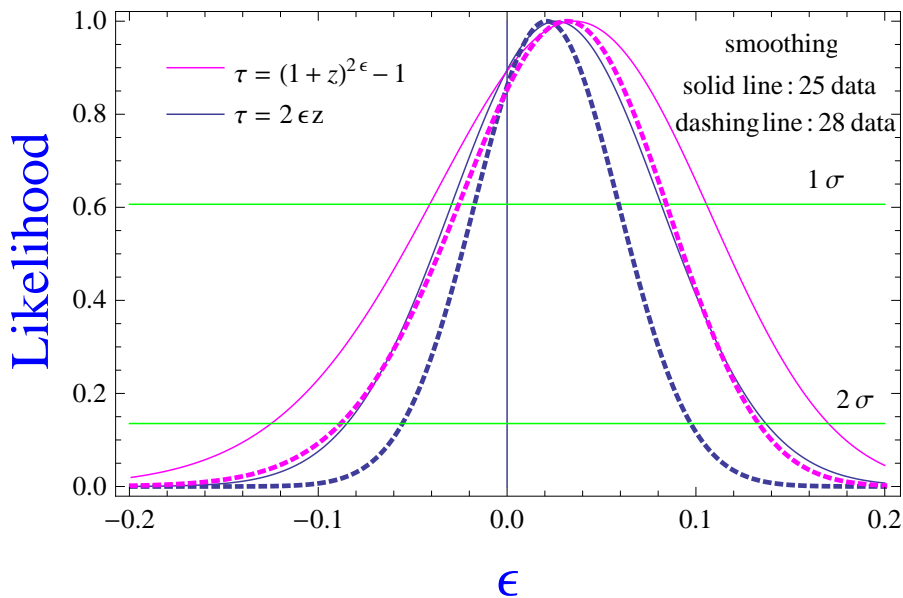}
\caption{Likelihood function for $\epsilon$ using smoothing method.
 }
\end{figure}

\section{Conclusion}
Until now, modern cosmology has discovered many interesting phenomena behind which there exists underlying physical
mechanisms. In principle, since the current astronomical observations are not precise enough to distinguish between different cosmological models, for example, various dark energy models are consistent with observations, we can explore
all the possibilities among which the true one exists. Though the matter component in the universe is so diluted, photons will get though a huge space to observers, photon conservation can be violated by simple astrophysical effects or by exotic physics. Amongst the former, the attenuation is due to interstellar dust, gas, plasmas and so on. More exotic sources of photon conservation violation involve a coupling of photons to particles beyond the standard model of particle physics. Therefore, the concept of cosmic opacity should be considered naturally. In
this Letter, we use the current observational Hubble parameter data which is opacity-free and SNe Ia observations which depends on cosmic opacity to test whether the universe is transparent. The results from three model-independent methods converge to a point that the effects of cosmic opacity is vanished. For future study on this problem, we think the wavelength is a considerable factor and more independent methods of testing cosmic opacity will confirm the conclusion.

\begin{table}
\begin{center}
\begin{tabular}{lcl}
\hline\hline
$z$ & $H(z)$ & $\sigma_{H}$ \\
\tableline
0.07 &69 &19.6\\
0.09 &69 & 12 \\
0.12 &68.6 &26.2\\
0.17& 83 & 8\\
0.179& 75 &4\\
0.199& 75 &5\\
0.2& 72.9& 29.6\\
0.24& 79.69 & 3.32\\
0.27& 77& 14\\
0.28 &88.8 &36.6\\
0.352 &83 &14\\
 0.4& 95& 17\\
0.43& 86.45& 3.27\\
0.44 &82.6 &7.8\\
0.48&97 &62\\
0.593& 104 &13\\
0.6  &87.9 &6.1\\
0.68 &92& 8\\
0.73 &97.3& 7\\
0.781 &105 &12\\
0.875 &125 &17\\
 0.88& 90& 40\\
 0.9& 117 & 23\\
 1.037 &154& 20\\
1.3& 168& 17\\
 1.43& 177& 18\\
 1.53& 140& 14\\
 1.75& 202& 40\\
\hline\hline
\end{tabular}
\end{center}
\caption{Current published observational Hubble parameter data ($km \cdot s^{-1} \cdot Mpc^{-1}$).
}\label{tab:Hz}
\end{table}

\textbf{\ Acknowledgments } This work was supported by the National Natural Science Foundation of
China under the Distinguished Young Scholar Grant 10825313,
the Ministry of Science and Technology National Basic Science Program (Project 973)
under Grant No.2012CB821804, the Fundamental Research Funds for the Central
Universities and Scientific Research Foundation of Beijing Normal University.

\end{document}